\documentclass[twocolumn,showpacs,preprintnumbers,amsmath,amssymb]{revtex4}

\usepackage{graphicx}
\usepackage{dcolumn}
\usepackage{bm}

\begin{document}

\title{{Inelastic scattering of broadband electron wave packets driven by an intense mid-infrared laser field}

\footnotetext{$^*$ E-mail: dichiara@mps.ohio-state.edu}}

\author{A. D. DiChiara$^*$, E. Sistrunk,  C. I. Blaga, U. B. Szafruga, P. Agostini and L. F. DiMauro}
\affiliation{ Department of Physics, The Ohio State University,
Columbus, Ohio 43210, USA\\}
\date{\today}

\begin{abstract}

{ 
Intense, 100 fs laser pulses at 3.2 and 3.6 $\mu$m are used to generate, by multi-photon ionization, 
broadband wave packets with up to 400 eV of kinetic energy and charge states up to Xe$^{+6}$. The multiple ionization pathways 
are well described by a white electron wave packet and {\it field-free} inelastic cross sections, averaged over the intensity-dependent energy distribution for (e,ne) electron impact ionization. The analysis also suggests a contribution from a 4$d$ core excitation in xenon.
}
\end{abstract}

 \maketitle

Ionization in a strong laser field can be described by a three-step process \cite{schafer,corkum} (Fig.1): production of photoelectrons, acceleration by the laser field, and (in)elastic recollisions or recombination with the parent ion resulting in above-threshold ionization (ATI), high harmonic generation (HHG) or multiple ionization \cite{agostini,walker,moshammer,feuerstein}. Each measureable conveys information on the intrinsic scattering event: a field-driven electron wave packet interacting with the core potential. Consequently, strong-field laser-driven scattering, or rescattering, is being vigorously pursued as a method for imaging molecules \cite{Lein}. Recently, {\it elastic} rescattering was shown to yield information similar to conventional electron beam experiments \cite{chen}, an established method for probing atomic and molecular structure \cite{zewail}. The link between conventional and laser-driven scattering is expressed multiplicatively with a collisional cross section and a calculated electron return distribution that represents a quantum wave packet \cite{nsiqrs}. While both methods are believed to be comparable, laser-driven elastic scattering presents an advantage because it is synchronized to the period of the laser field. Consequently, it is considered a potential method for probing structure with sub-femtosecond resolution \cite{niikura,blaga}. However, as with all novel methods, laser-driven scattering requires more rigorous treatment that must be obtained through broader analysis. Currently, efforts for developing laser-driven scattering have relied on the analysis of HHG and ATI. In this letter we investigate the third recollision channel, {\it inelastic} scattering, by observing multiple ionization. Our experiment offers a new comprehensive basis for assessing laser-driven scattering and emphasizes the use of inelastic events as a probe of complex electronic structure. 

Laser-driven inelastic scattering can result in non-sequential ionization (NSI) through (e,2e) \cite{walker,moshammer} or higher (e,ne) processes \cite{feuerstein}, or excitation followed by field ionization \cite{ivanov}. Most studies hitherto have been performed at near-infrared (NIR) wavelengths, ($\lambda\leq$ 1 $\mu$m). These investigations are limited to measurements of double NSI since depletion of the neutral ground state restricts the maximum intensity, and thus the return energy, to the (e,2e) inelastic channel. This limitation holds for all neutral atoms ionized by a NIR field. In addition, extraction of the inelastic NSI channel is masked by the presences of photon-assisted processes. In helium for example, semi-classical analyses \cite{ivanov,nsiqrs} show contributions from both (e,2e) inelastic and field-ionized excited states. For higher-Z atoms, multiphoton \cite{rudati} and higher-order sequential \cite{chin, sasi} processes complicate the analysis of the observed ion yields. Critically, the ability to exploit laser-driven scattering as a tool relies on the comprehensive understanding of all these experiments and the accuracy of a field-free description of rescattering in a strong laser field. 

In this Letter, intense long wavelength pulses (3.6 and 3.2 $\mu$m) result in a $\lambda^2$ increase in the electron's rescattering energy \cite{colosimo} allowing the extension of NSI studies to (e,ne) processes for n=2-6. The corresponding observations are both simplified and more comprehensive because the field strength can be kept low enough to ionize only the neutral or excited singly charged ion and at the same time open a plethora of (e,ne) channels. We will show that our results can be interpreted by using known electron impact cross sections \cite{achenbach2, muller}. Thus, the link between NSI and laser-driven recollision is addressed over a broader, more significant, energy range. Moreover, at such impact energies, both inner-shell and valence ionization pathways are allowed in xenon \cite{achenbach}. This, by comparison, should illuminate the nature of laser-driven multiple ionization.

\begin{figure}
	\centering
   	\includegraphics[width=0.50\textwidth]{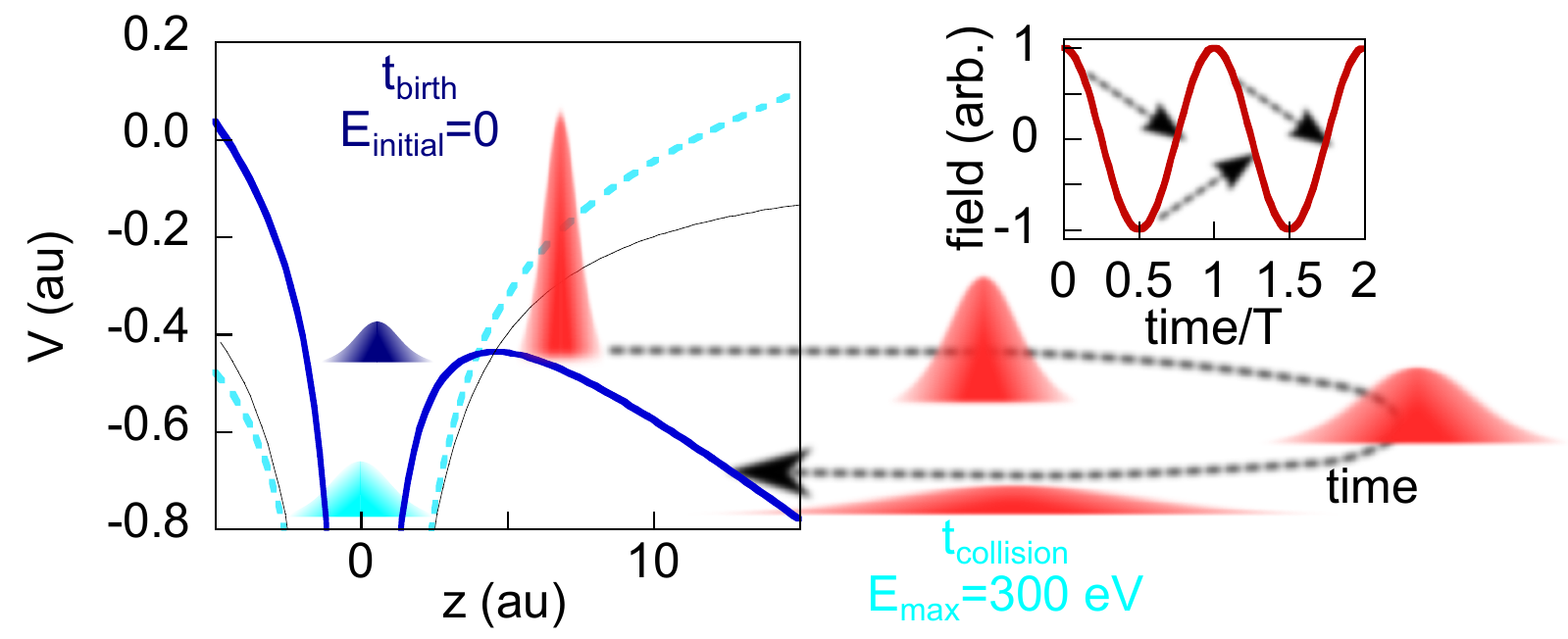}
	\caption{Diagram of the laser suppressed Coulomb potentials with the strong-field photoionization sequence illustrated: ionization, acceleration in the field with lateral expansion, and recollsion with a distorted potential. The Coulomb + laser potential of the neutral, dark blue solid line, and ion, light blue dashed line, are shown at a field intensity of 90 TW/cm$^2$ for the phase of birth that yields the maximum return energy of 3.2$U_P$. For reference the bare ionic Coulomb potential is shown by a thin black solid line.}
	\label{fig:Fig1aug2.pdf}
\end{figure}

We will show that laser-driven scattering is comparable to conventional electron impact ionization broadened by the nearly white returning wave packet that, according to classical mechanics, spans from 0 to 3.2$U_P$ \cite{schafer, corkum} where $U_P$ is the ponderomotive potential given by $U_P$[eV]=93$\cdot$Intenstity[PW/cm$^2]\cdot\lambda^2[\mu$m]. For comparison, conventional experiments are performed with nearly monochromatic beams. A large $U_P$ and modest binding potential, $I_P$, imply tunnel ionization since the Keldysh parameter \cite{keldysh}, $\gamma=\sqrt{I_P/2U_P}$, is less than unity. Our mid-infrared (MIR), 100 fs, laser is tuned to either 3.6 or 3.2 $\mu$m and provides: 25 eV $> U_P >$120 eV, $\gamma \leq 0.5$, and electron wave packets with up to 400 eV of rescattering energy.

The laser is described elsewhere \cite{IEEE} and consists of a difference frequency generation (DFG) amplifier fed by two independent laser systems. The 100 fs pump beam is generated by a Ti:Sapphire chirped pulse amplifier and the 16 ps signal beam is generated by a regeneratively amplified Nd:YLF laser \cite{saed}. DFG is performed in a KTA crystal and the idler beam has a central wavelength tunable to 3.6 or 3.2 $\mu$m, a 100 fs pulse duration and a peak power exceeding 1 GW. The beam is focused into a time of flight mass spectrometer with a base pressure of $10^{-9}$ torr and $10^9-10^{12}$ cm$^{-3}$ target densities. The mass resolution of the spectrometer is $\Delta m/m$$\approx$$0.3\%$ and fully resolves all xenon isotopes up to the highest charge state observed. An intensity calibration, accurate to within $\pm 10 \%$, was performed by measuring photoelectron spectra that have a characteristic change in slope at 2$U_P$ \cite{colosimo}.

\begin{figure}
	\centering
	\includegraphics[width=0.48\textwidth]{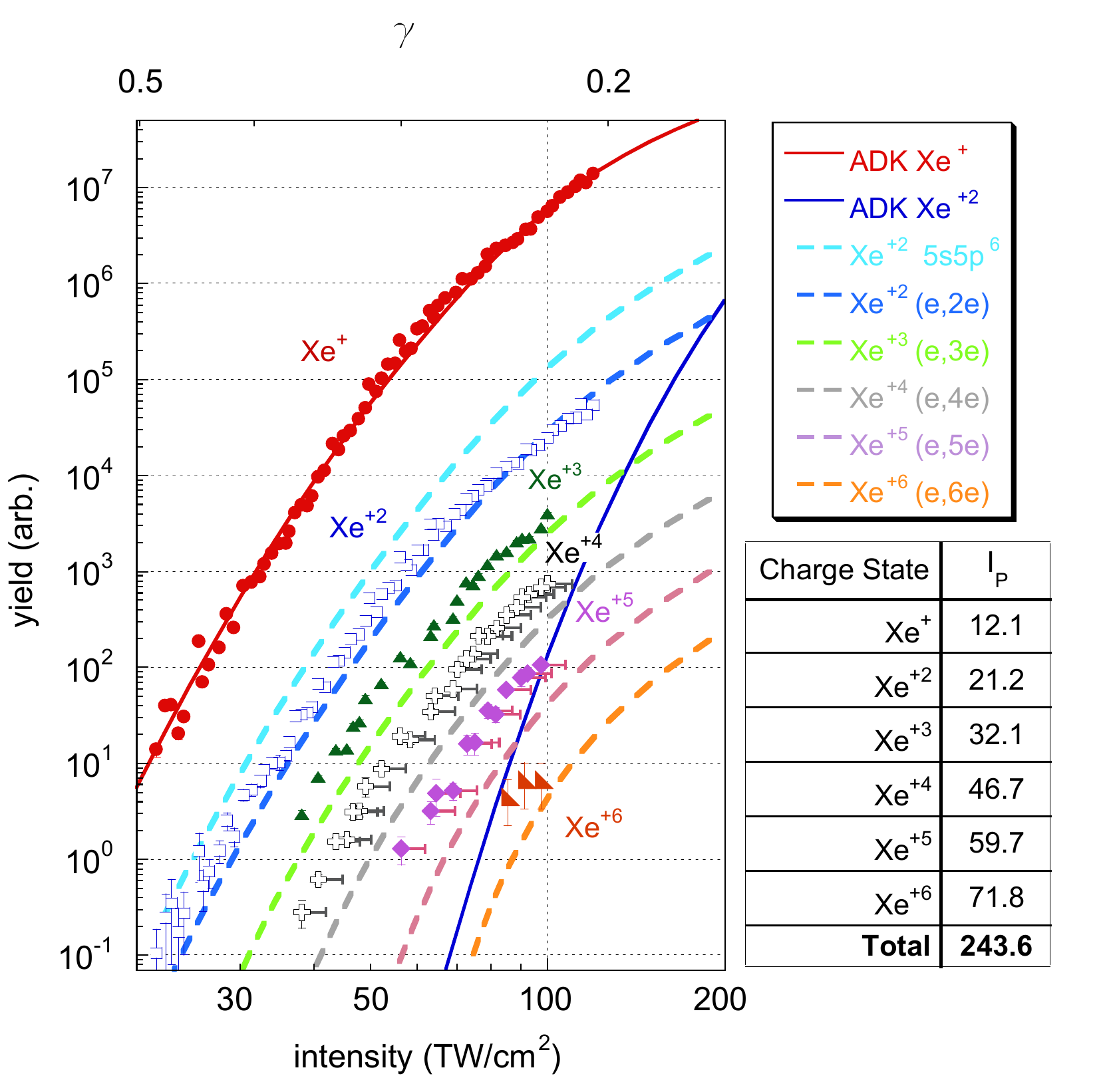}
	\caption{Ionization yields of xenon at 3.6 $\mu$m are shown with various symbols. The solid lines are calculated with ADK theory (see text). Top horizontal axis is the Keldysh parameter. The vertical error bars are given by Poisson statistics and are, in general, smaller than the size of the symbols. Horizontal error bars show the 10$\%$ uncertainity in our intensity estimate. The calculated nonsequential yields are shown with dashed lines and, for reference, the ionization potentials for Xe$^+$-Xe$^{+6}$ are given.} 
\end{figure}

Figure 2 shows the measured ion yields of xenon at 3.6 $\mu$m where the highest charge state observed is Xe$^{+6}$. At 3.2 $\mu$m, not shown, the highest charge state observed is Xe$^{+5}$. To verify that the ground state of the neutral is directly tied to the generation of higher charge states the sequential yield of Xe$^{+2}$ was calculated using the Ammosov, Delone and Krainov (ADK) \cite{ADK} tunneling rate. Whereas the Xe$^+$ measured yield is in close agreement with ADK for both 3.6 and 3.2 $\mu$m the predicted sequential yield of Xe$^{+2}$ underestimates the observed yield. In fact, Xe$^{+5}$ is produced in greater abundance. Non-sequential behavior is well known in the NIR domain and should not be different in the MIR. However, at NIR wavelengths in xenon \cite{chin, rudati} non-sequential double ionization results from the multiphoton characteristics, {\emph e.g.} $\gamma>1$, of the ionization sequence and excitation. In the current study, observation of charge states up to Xe$^{+6}$ is evidence of an extreme non-sequential process where over 700 MIR photons are coupled to neutral xenon liberating 6 electrons bound by a net potential of 243.6 eV. We have calculated non-sequential yields, shown with dashed lines in Fig. 2 and discussed below, with a nearly white wave packet and (e,ne) cross sections and found excellent agreement.

Our analysis begins by defining inelastic branching ratios as Xe$^{+n}$/Xe$^{+}$, $n>$ 1, and assuming that the yield of the $n^{th}$ charge state is proportional to photoionization of the neutral. Since the returning electron flux is proportional to the ionization rate, our branching ratios, see Figs. 3(c) and 3(d), are effective, energy averaged cross sections for laser-driven multiple ionization. For comparison, the known experimental cross sections measured with electron guns \cite{muller, achenbach2}, are shown in Fig. 3(b). Experimental data is not, to the best of our knowledge, available for both quintuple ionization and 5s5p$^6$ excitation of Xe$^+$ so we have used the Lotz formula \cite{lotz} to scale measured cross sections of Xe$^+$. As an example, we observe a threshold response in our branching ratios that is similar to what occurs in the measured cross sections. This is most clearly seen in our experiment at 3.2 ${\mu}$m where the ponderomotive potential is 25$\%$ lower than at 3.6 ${\mu}$m. 

\begin{figure}
	\centering
	\includegraphics[width=0.47\textwidth]{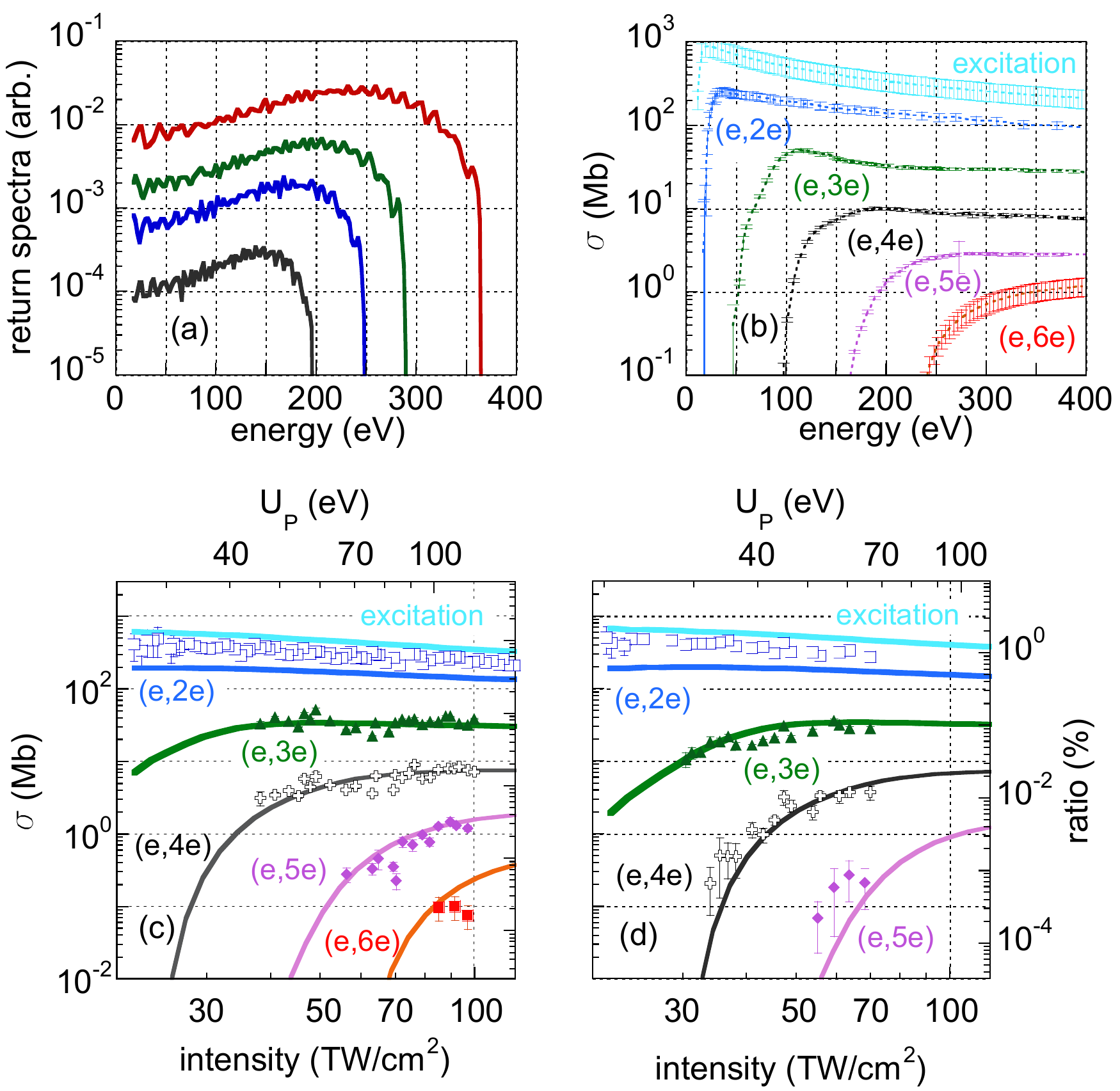}
	\caption{The returning photoelectron distributions are shown for 100 (red), 80 (green), 64 (blue) and 51 (black) TW/cm$^2$ in panel (a). For reference the collisional excitation (5s5p$^6$) and ionization cross sections used in our calculation are shown in panel (b), see text for details. The error bars for (e,2e)-(e,5e) are taken directly from refs. \cite{achenbach2, muller}. For 5s5p$^6$ excitation and (e,6e) a Lotz formula is assumed with a minimum error of ±25$\%$. Panels (b) and (c) show the results of Eqn. 1 for 3.6 and 3.2 $\mu$m respectively, where the intensity axis is determined by the calculation, and the vertical overlap is referenced against the full NSI calculation shown in Fig. 2.}
\end{figure}
 
As already mentioned, the spectral width of the returning electron distribution is known to be very broad, due to the distribution of ionization times \cite{nsiqrs}. In addition, laser experiments are necessarily integrated over the spatial and temporal intensity distributions of the  focussed beam. Hence, we define an effective, energy-averaged \cite{atomicbook}, cross section, $\tilde{\sigma}(I)$ as:
\begin{equation}
\label{seff}
\tilde{\sigma}(I)=\frac{\int dE' \sigma(E') W_P(E'(I))}{\int dE' W_P(E'(I))},
\end{equation}
where $W_P$ is the net return distribution, and $\sigma(E)$ is the experimental cross section.
Consequently, the effective cross section is a function of the peak intensity $I$.

To calculate $W_P$ the laser pulse is sampled in time increments of 1 au and at each point a 1-D trajectory is calculated by solving $\ddot{x}=-E(t)$; only trajectories that return to the core are retained. The initial conditions are consistent with the treatment in \cite{ivanov, sasi}. Each trajectory is weighted by the ADK yield to account for tunnel ionization and ground state depletion.
The ballistic expansion of a tunnel ionized wave packet \cite{delone and krainov} is taken into account to ensure that each classical trajectory represents a wave packet with the correct area at recollision. Finally, the calculation is summed for individual volume elements of a Gaussian focus. 

Examples of the return distributions are shown in Fig. 3(a). Our classical representation of the returning wave packet strongly resembles the quantum mechanical version calculated in \cite{nsiqrs}. In general each distribution has a width of approximately 1.5$U_P$, increases slowly by a factor of 3 to a maximum at around 2$U_P$ and then decreases rapidly at a maximum return energy of 3.2$U_P$ \cite{schafer}.

The average recolliding wave packet has an $e^{-1}$ width of more than a nanometer. For reference, the (e,2e) cross section shown in Fig. 3(b) has a width of about 1  Angstrom. Therefore, we need to include the dependence of the ionization probability on the impact parameter \cite{excqrs,ivanov} since this determines the effective current density of the return distribution and quantifies the observed yield from the effective cross section Eqn. (\ref{seff}):
\begin{equation}
\label{proba}
P(b,I)=\tilde{\sigma}(I)\frac{\exp(b^2/a_0^2)}{\pi a_0^2},
\end{equation}
where $a_0=\sqrt{2/\Delta E}$ and b is the impact parameter.

Our approach allows two analyses of laser-driven scattering: the first investigates the role of an energetically broad wave packet and relates a laser-driven {\it measured} branching ratio to a field-free {\it measured} electron impact cross section. The second accounts for the lateral expansion of the wave packet and quantifies the observed yield. For both cases we use the largest impact excitation cross section in xenon, 5$s$5$p^6$, and (e,2e) impact ionization to account for Xe$^{+2}$ yield, and (e,3e)-(e,6e) impact ionization cross sections for Xe$^{+3}$-Xe$^{+6}$, respectively. Xe$^{+*}$ has a binding potential of 9.9 eV and is assumed to ionize with unity probability. In Fig. 2, the yields, and in Figs. 3(c) and 3(d), the branching ratios, from excitation and (e,2e) ionization are plotted separately to show the different contributions from each channel. We see that for both analyses, and for both 3.6 and 3.2 $\mu$m laser fields, the yield predicted from excitation alone overestimates our experimental results while the observed multiple ionization channels agree with our assumption of a direct (e,ne) ionization event. We argue that excitation ionization in NSI may be beyond a semi-classical analysis because: first, excitation cross sections are typically inferred through a fitting procedure \cite{excqrs} and may have large errors. Second, recent work \cite{schafer2} suggests that higher lying excited states are coupled to the field and may not be accurately described by field-free cross sections. 

Experimental and calculated NSI yields for Xe$^{+3}$ and higher agree within a $10\%$ uncertainty in our intensity estimate and our overall experimental error (Fig.2). In addition, the excellent agreement between observed branching ratios and effective cross sections for channels (e,ne) (n $\geq 3$) (Fig.3 (c) and (d)) justifies the correspondence between laser-driven inelastic scattering and field-free impact ionization and emphasizes the key role of the energetically broad recolliding wave packet. Moreover, it justifies retaining only channels that couple higher charge states to photoionization of the neutral. Finally, experiments performed with krypton for processes up to (e,5e) were found to be consistent with our  interpretation.

\begin{figure}
	\centering
	\includegraphics[width=0.50\textwidth]{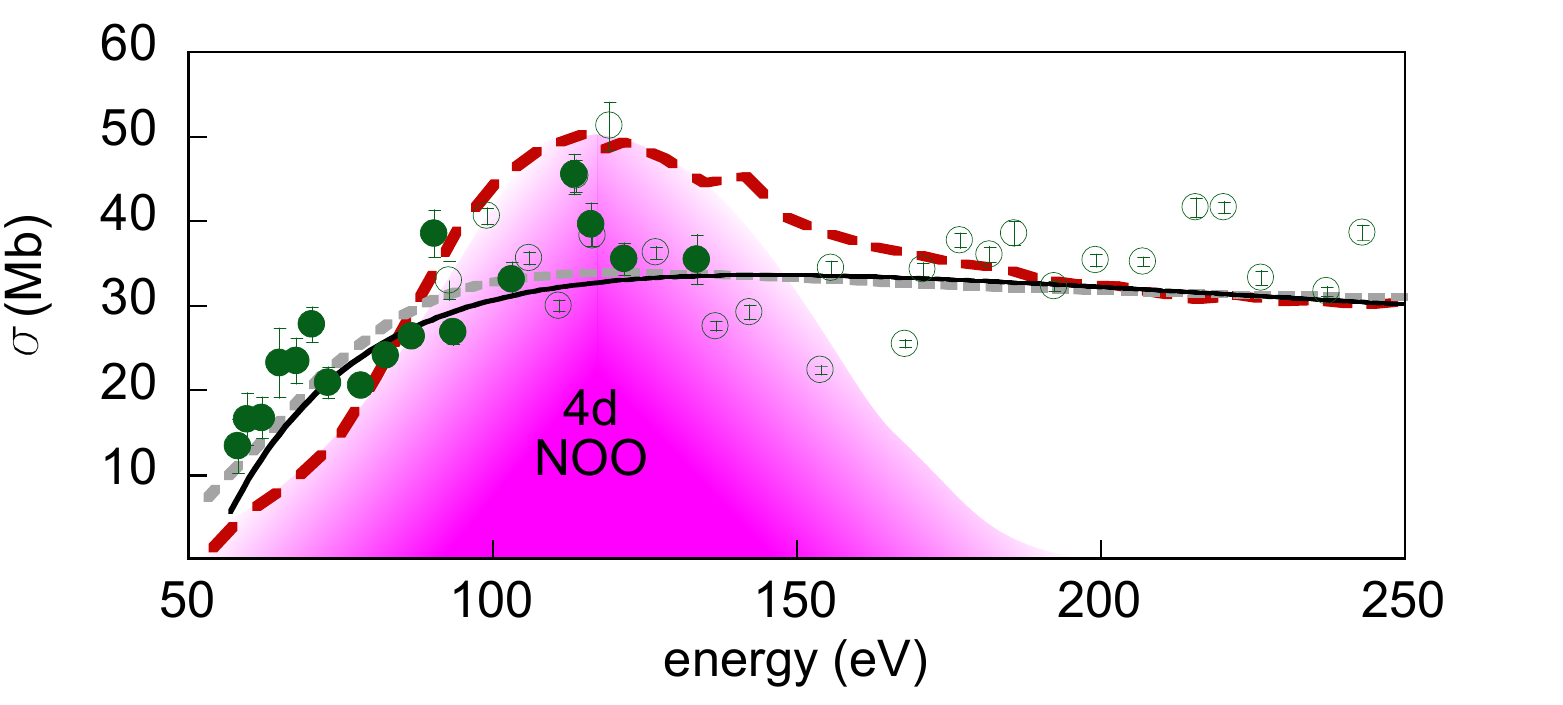}
	\caption{ The measured (e,3e) cross section from \cite{achenbach}, red long dash, effective cross section from Eqn. 1, grey short dash, and the (e,3e) direct ionization channel, thin black line, obtained with a Lotz fit \cite{lotz} are shown to highlight the feature between 100-150 eV. Photoionization of a 4$d$ electron is indicated by the purple shaded region, see text for details. Measured +3/+ branching ratios for 3.2 ${\mu}$m, green solid circle, and 3.6 ${\mu}$m, green open circle, are shown on an energy scale determined by the peak of the return distribution that occurs at 2$U_P$.}
\end{figure}

A qualitative argument that justifies our field-free interpretation appears in Fig. 1 which shows the Stark ionic potentials for the maximum return energy of 3.2$U_P$. Because the recolliding electron returns near the zero of the laser field the Coulomb potential suffers only a small distortion. We also note that even at the field maximum the region of the potential near $I_P(ion)$, light blue Gaussian in Fig. 1, is not significantly distorted.  The success of the field-free cross sections has thus, a simple explanation in the three-step semi-classical model.

Finally, the high ponderomotive energy available at long wavelengths opens more complicated ionization pathways. For example, (e,3e) ionization may involve an Auger (NOO) channel, or giant resonance, where the primary electron is removed from the 4$d$ sub-shell \cite{achenbach, muller}. Achenbach {\it et al.} \cite{achenbach} supported this argument with a comparison of their electron impact ionization data with synchrotron photoionization experiments \cite{haensel}. Their data along with the highlighted photoionization feature is shown in Fig. 4. For reference, both the effective, energy averaged cross section from Eqn. 1 and the (e,3e) direct ionization cross section are shown in Fig. 4. Our (e,3e) data is consistent with the feature identified in \cite{achenbach} and has a small hump around $100-150$ eV that suggests inner-shell ionization is also occurring in the laser-driven experiment. The role of the energetically broad wave packet here is that it averages out finer features.
 
In conclusion we have shown that multiple ionization occurs from laser-driven scattering with a MIR laser. The maximum observed charge state is Xe$^{+6}$ which is shown to occur through a direct field-free (e,6e) channel. The link between field-free impact ionization and laser-driven NSI is shown to be the broadband returning wave packet that effectively averages the cross sections. An inner-shell Auger process is suggested by the data and illustrates the effect of energy averaging. Our analysis of multiple ionization will play a critical role in helping to establish laser-driven scattering as a tool and an application of strong-field science.

L.F.D. acknowledges support from the Hagenlocker chair. This work was supported by the NSF under grant PHY-1004778.


\end{document}